\documentclass[12pt]{article}

\usepackage{amsmath}
\numberwithin{equation}{section}

\evensidemargin \oddsidemargin

\begin{document}

\makeatletter

\renewcommand{\thefootnote}{\fnsymbol{footnote}}
\newcommand{\beq}{\begin{equation}}
\newcommand{\eeq}{\end{equation}}
\newcommand{\bea}{\begin{eqnarray}}
\newcommand{\eea}{\end{eqnarray}}
\newcommand{\nn}{\nonumber\\}
\newcommand{\hs}[1]{\hspace{#1}}
\newcommand{\vs}[1]{\vspace{#1}}
\newcommand{\Half}{\frac{1}{2}}
\newcommand{\p}{\partial}
\newcommand{\ol}{\overline}
\newcommand{\wt}[1]{\widetilde{#1}}
\newcommand{\ap}{\alpha'}
\newcommand{\bra}[1]{\left\langle  #1 \right\vert }
\newcommand{\ket}[1]{\left\vert #1 \right\rangle }
\newcommand{\vev}[1]{\left\langle  #1 \right\rangle }

\newcommand{\ul}[1]{\underline{#1}}
\newcommand{\tr}{\mbox{tr}}
\newcommand{\ishibashi}[1]{\left\vert #1 \right\rangle\rangle }

\makeatother
 
\def\a{\alpha} 
\def\b{\beta} 
\def\c{\gamma}  
\def\d{\delta} 
\def\e{\epsilon}           
\def\f{\phi}               
\def\g{\gamma} 
\def\h{\eta}    
\def\i{\iota} 
\def\j{\psi} 
\def\k{\kappa}                    
\def\l{\lambda} 
\def\m{\mu} 
\def\n{\nu} 
\def\o{\omega}  \def\w{\omega} 
\def\p{\pi}                
\def\q{\theta}  \def\th{\theta}                  
\def\r{\rho}                                     
\def\s{\sigma}                                   
\def\t{\tau} 
\def\u{\upsilon} 
\def\x{\xi} 
\def\z{\zeta} 
\def\D{\Delta} 
\def\F{\Phi} 
\def\G{\Gamma} 
\def\J{\Psi} 
\def\L{\Lambda} 
\def\O{\Omega}  \def\W{\Omega} 
\def\P{\Pi} 
\def\Q{\Theta} 
\def\U{\Upsilon} 
\def\X{\Xi} 
\def\del{\partial}              
 
 
\def\ca{{\cal A}} 
\def\cb{{\cal B}} 
\def\cc{{\cal C}} 
\def\cd{{\cal D}} 
\def\ce{{\cal E}} 
\def\cf{{\cal F}} 
\def\cg{{\cal G}} 
\def\ch{{\cal H}} 
\def\ci{{\cal I}} 
\def\cj{{\cal J}} 
\def\ck{{\cal K}} 
\def\cl{{\cal L}} 
\def\cm{{\cal M}} 
\def\cn{{\cal N}} 
\def\co{{\cal O}} 
\def\cp{{\cal P}} 
\def\cq{{\cal Q}} 
\def\car{{\cal R}} 
\def\cs{{\cal S}} 
\def\ct{{\cal T}} 
\def\cu{{\cal U}} 
\def\cv{{\cal V}} 
\def\cw{{\cal W}} 
\def\cx{{\cal X}} 
\def\cy{{\cal Y}} 
\def\cz{{\cal Z}}

 
\def\bop#1{\setbox0=\hbox{$#1M$}\mkern1.5mu 
        \vbox{\hrule height0pt depth.04\ht0 
        \hbox{\vrule width.04\ht0 height.9\ht0 \kern.9\ht0 
        \vrule width.04\ht0}\hrule height.04\ht0}\mkern1.5mu} 
\def\Box{{\mathpalette\bop{}}}                        
\def\pa{\partial}                              
\def\de{\nabla}                                       
\def\dell{\bigtriangledown} 
\def\su{\sum}                                         
\def\pr{\prod}                                        
\def\iff{\leftrightarrow}                      
\def\conj{{\hbox{\large *}}} 
\def\lconj{{\hbox{\footnotesize *}}}          
\def\dg{\sp\dagger} 
\def\ddg{\sp\ddagger} 
 
\def\>{\rangle} 
 
\def\<{\langle} 
\def\Dsl{D \hskip-.6em \raise1pt\hbox{$ / $ } } 

\def\phil{@{\extracolsep{\fill}}} 
\def\unphil{@{\extracolsep{\tabcolsep}}}

\def\to{\rightarrow} 
\def\pa{\partial} 
\def\del{\nabla} 
\def\delbar{\bar{\nabla}} 
\def\xx{\times} 
\def\ab{\bar{a}} 
\def\bb{\bar{b}} 
\def\cb{\bar{c}} 
\def\db{\bar{d}} 
\def\eb{\bar{e}} 
\def\fb{\bar{f}}  \def\Fb{\bar{\F}} 
\def\kb{\bar{k}}  \def\Kb{\bar{K}} 
\def\lb{\bar{l}}  \def\Lb{\bar{L}}  
\def\mb{\bar{m}}  \def\Mb{\bar{M}} 
\def\nb{\bar{n}}  \def\Tb{\bar{T}}  
\def\zb{\bar{z}}  \def\Gb{\bar{G}}  
\def\wb{\bar{w}}  \def\Jb{\bar{J}} 
\def\pb{\bar{p}}   
\def\Ib{\bar{I}} 
\def\Pbb{\bar{\P}}  
\def\jbb{\bar{\j}} 
\def\qbb{\bar{\q}} 
\def\Sbb{\bar{\S}} 
\def\Pbb{\bar{\P}} 
\def\dbb{\bar{\delta}} 
\def\kbb{\bar{\kappa}}  
\def\ebb{\bar{\epsilon}} 
\def\ua{\underline{a}} 
\def\ub{\underline{b}} 
\def\uc{\underline{c}} 
\def\ud{\underline{d}} 
\def\ue{\underline{e}} 
\def\uua{\underline{\a}} 
\def\uub{\underline{\b}} 
\def\uuc{\underline{\c}} 
\def\uud{\underline{\d}} 
\def\uue{\underline{\e}} 
\def\uug{\underline{\g}} 
\def\hL{\hat{L}}  
\def\hM{\hat{M}} 
\def\hj{\hat{\j}}  
\def\hf{\hat{\f}} 
\def\hMb{\bar{\hat{M}}} 
\def\hfb{\bar{\hat{\f}}} 
\def\hP{\hat{\Psi}} 
\def\dif{\partial} 
\def\difb{\bar{\partial}} \def\dbar{\bar{\partial}} 
\def\pab{\bar{\pa}} 
\def\nonu{\nonumber \\{}} 
\def\half{{1 \over 2}} 
\def\bfx{{\bf x}} 
\def\bfy{{\bf y}} 
\def\bfk{{\bf k}} 
\def\bfl{{\bf l}} 
\def\vp{{\vec p}} 
\def\vx{{\vec x}} 
 
\def\cn{${\bf C}/{\bf Z}_N\ $} 
\def\ppten{{$PP_{10}$ }} 
\def\ppsix{{$PP_{6}\times R^4$ }} 

\begin{titlepage}

\renewcommand{\thefootnote}{\fnsymbol{footnote}}

\hfill\parbox{4cm}{ hep-th/0404081 \\KAIST-TH 2004/03}

\vspace{35mm}
\baselineskip 9mm
\begin{center}
{\LARGE \bf A note on D-brane Interactions\\
in the IIA Plane-Wave Background}
\end{center}

\baselineskip 6mm
\vspace{10mm}
\begin{center}
Yeonjung Kim,$^a, ^b$\footnote{\tt geni@muon.kaist.ac.kr} 
and Jaemo Park$^c$\footnote{\tt jaemo@physics.postech.ac.kr} 
\\[5mm] 
{\sl $^a$Department of Physics, KAIST, 
Taejon 305-701, Korea \\ 
$^b$ R \& D Division, Hynix Semiconductor Inc., \\
San 136-1 Ami-ri Bubal-eub Icheon, 467-701, Korea  \\
$^c$ Department of Physics, POSTECH, Pohang 790-784,
Korea \\} 
\end{center}

\vspace{20mm}

\thispagestyle{empty}

\vfill
\begin{center}
{\bf Abstract}
\end{center}
We compute the D-brane tensions in the type IIA plane-wave background by
comparing the interaction potential between widely separated D-branes in
string theory with the supergravity mode exchange between the D-branes.
We found that the D-brane tensions and RR charges in the plane-wave background are the same
as those in the flat space. Also we work out the stringy halo behavior\cite{Johnson}
of the $DpD\bar{p}$ spacelike branes and find the explicit dependence
on the light cone separation $r^+$. This suggests that the detailed
tachyon dynamics for the spacelike $DpD\bar{p}$ branes are different from those 
in the flat space case. We also discuss specific features of the exchange amplitudes
in relation to the geometric properties of the IIA plane wave background. When branes are
located at focal points, full or partial restoration of the translation invariance occurs
and the amplitudes are similar to those in the flat space. 

\vspace{20mm}
\end{titlepage}

\baselineskip 6.5mm
\renewcommand{\thefootnote}{\arabic{footnote}}
\setcounter{footnote}{0}

\section{Introduction}

In a series of papers\cite{SPH},\cite{SSY},\cite{HS},\cite{HS2}
,\cite{HS3},\cite{Yoshida},\cite{Me}, \cite{gan}  simple type IIA string theory on the plane wave
background have been studied, parallel to the development of the
IIB string theory on the plane wave background \cite{Metsaev}
,\cite{Tseytlin}, \cite{Gaberdiel}
,\cite{Bain},\cite{Dabholkar}, \cite{cot}.
The background for the string theory is obtained by compactifying the 11-
dimensional plane wave background on a circle and taking the small radius limit.
By working out the details of the resulting string theory one can obtain the useful
information about the M-theory on the 11-dimensional plane wave background.
However the resulting string theory in itself has many nice features and is worthy of
the detailed study. It admits light cone gauge where the string theory spectrum is
that of the free massive theory, which is similar to the IIB case.
The world sheet enjoys (4,4) world sheet supersymmetry and the supersymmetry
commutes with the Hamiltonian so that all members of the same supermultiplet has
the same mass, which is different from the IIB theory.
The various $1/2$ BPS D-brane states were analyzed both in the light cone gauge
and in the covariant formalism\cite{SPH},\cite{SSY}.
Finally the boundary state formalism has developed in \cite{SSY},\cite{Me}. And the usual
open-closed string channel duality was shown to hold for the supersymmetric
D-brane configurations. In this note, we  study the
interactions of the D-branes and the anti-D-branes in the Type IIA
plane wave background based on the boundary state formalism. 
As  applications of this development, we work out the D-brane tensions,
Ramond-Ramond (RR) charges and the stringy halo\cite{Banks, Johnson}
 of the D-branes in the IIA plane
wave background. 

First we work out the open string partition function in the presence of widely
separated D-branes. By taking a suitable limit, we find the contribution from the
supergravity modes. By comparing this string theory computation with the corresponding
supergravity calculation, we find that the supersymmetric D-brane tensions and RR charges are
the same as those in the flat space.
Similar result was obtained for the various IIB plane wave backgrounds \cite{Dabholkar}.
In fact, we closely follow their approach in this paper.
As explained in \cite{Dabholkar}, for the timelike branes, the result is more or
less trivial. The interaction between a brane and an anti-brane is given by
the overlap $\bra{D\bar{p}} \D \ket{Dp}$ where $\D$ is the closed string propagator
and $\ket{Dp}, \ket{D\bar{p}}$ are the corresponding boundary states.
Due to the boundary condition, they satisfy $\hat{p^+}\ket{Dp} = \hat{p^+}\ket{D\bar{p}}$ = 0
where $\hat{p^+}$ is the light cone momentum operator. This condition projects the
closed string propagator to the $\hat{p^+} = 0$ subspace and these states propagate as
in the flat space.

However it is nontrivial to compute the tension of the space-like D-branes. 
Actual calculation shows that the supersymmetric D-brane tensions and RR charges are 
the same as those in
flat space irrespective of their position in the flat space. In view
of this, one might think that space-like D-branes in the IIA plane
wave background would behave in the same way as the timelike branes do.
But another calculation shows that this is not the case. We ask at
which distance 
$DpD\bar{p}$ branes would develop the divergence of the amplitude,
which indicates the instability
associated with the decay into the closed string channel. In the flat
space case, $DpD\bar{p}$ branes develop the divergence at the
interbrane distance $X_H\equiv \sqrt{2\pi^2 \alpha'}$. 
This was interpreted as D-branes having a stringy halo. This was also
discussed in \cite{Johnson} in the context of $D(-1)$ instantons in
the IIB plane wave background. We will see that the space-like
$DpD\bar{p}$ branes in the IIA plane wave background 
would develop the stringy halo but the halo
itself depends on the mass parameter, which again has the nontrivial 
dependence on the light-cone separation $r^+$. Thus we confirm the
different behavior of the stringy halo for the various spacelike
D-branes in the IIA plane wave background than that of the spacelike 
D-branes in the flat space. This was observed for $D(-1)$ brane in
the IIB plane wave background and we confirm that spacelike D-branes 
in the IIA plane wave background have the similar characters. 
On the other hand, timelike  branes in the IIA plane wave background 
has the same stringy halo behavior as the D-branes in the flat space
since in the boundary state formalism the closed string propagator
projects to the $p^+=0$ subspace.
Thus one can say that the timelike branes and the spacelike branes 
in the IIA plane wave background would behave in the similar way at
the large interbrane distance but they behave differently at the short 
interbrane distance. In view of this, the fact that spacelike branes 
in the plane wave background have the same tension as that of the
D-branes in the flat space is quite nontrivial. Also this would
indicate that the spacelike branes are the good probes to study the 
different behaviors of D-branes in the IIA plane wave background than 
that of D-branes in the flat space.
Thus the next logical step would be to study the various interaction
terms of the D-branes using either boundary state formalism or 
using the Dirac-Born-Infeld type action of the D-branes in the IIA 
plane wave background. This is also needed to understand the tachyon
dynamics in the IIA plane wave background.

Finally we discuss the specific features of the integrated amplitudes. 
Motivated by the similar work in the Type IIB side \cite{skenderis}, we can deduce 
the many features of the integrated amplitudes from the geometrical properties
of the Type IIA plane wave. Several features are traced to the lack of translational 
invariances. For specific values of $r^+$ the translational invariances are 
fully or partially restored, which is due to the focusing of the geodesics. 
The reinstated translational invariance can be seen from the world sheet calculation
as well.

The content of this note is as follows. 
In section 2, we present the string theory calculation for the interaction potential 
between two D-branes. Then we discuss the case where the interbrane
distance is large so that we can compare with the tensions obtained 
from the field theory calculation. This is followed by the
consideration of the short distance limit where the stringy halo develops.
In section 3 we carry out the supergravity analysis. In section 4, followed is 
the discussion of 
the integrated amplitudes in relation to the geometric properties of the Type IIA plane
wave background. In the
appendix, we provide the relevant information of the supergravity bulk action relevant
for the calculation in the text.

\section{String Theory Calculation}
The IIA plane wave background of our interest is given by 
\beq
ds^2 = -2dX^+dX^--A(x^I)(dX^+)^2+\sum_{i=1}^{8}dX^IdX^I,
\eeq
\beq
F_{+123} =\mu , F_{+4} =\frac{\mu}{3}
\eeq
and 
\beq
A(x^I)=\sum_{i=1}^{4}\frac{\mu^2}{9}(X^{i})^2+\sum_{i'=5}^{8}\frac{\mu^2}{36}(X^{i'})^2
\eeq
where $X^\pm=\frac{1}{\sqrt{2}}(X^0\pm X^9)$.
With the presence of the Ramond-Ramond background, the above plane wave background
has $SO(3)\times SO(4)$ where $SO(3)$ acts on
1,2,3 directions and $SO(4)$ acts on 5,6,7,8 directions.
We use the convention that unprimed coordinates denote 1,2,3,4 directions
while primed coordinates denote 5,6,7,8 directions.
For the worldsheet coordinates we use $\pa_\pm=\Half(\pa_\tau\pm\pa_\sigma)$.

The worldsheet action for the closed string theory is given by 
\bea
S_{LC} & = &- \frac{1}{4\pi\ap}\int d^2\sigma (-\pa_\mu X^+\pa^\mu X^-
+\pa_\mu X^I\pa^\mu X^I + \frac{m^2}{9}\sum_{i=1}^{4}X^iX^i
+\frac{m^2}{36}\sum_{i'=1}^{4}X^{i'}X^{i'}
\nn
 & & + 2 \sum_{b=\pm} (-i\psi_b^1\pa_{+}\psi_b^1-i\psi_b^2\pa_{-}\psi_b^2)
+\frac{2im}{3}\psi_{+}^2\gamma^4\psi_{-}^1
-\frac{im}{3}\psi_{-}^2\gamma^4\psi_{+}^1)
\label{f1}
\eea
where $m=\ap p^+\mu$ and 
$\gamma^I$ are $8\times 8$ matrices satisfying $\{\gamma^I, \gamma^J\}=\delta^{IJ}$.
The sign of subscript $\psi_{\pm}^A$ denotes the eigenvalue of $\gamma^{1234}$ while the 
superscript $A=1,2$ denotes the eigenvalue of $\gamma^9$.
The theory has two supermultiplets $(X^i,\psi^1_-,\psi^2_+)$, $(X^{i'},\psi^1_+,\psi^2_-)$
of $(4,4)$ worldsheet supersymmetry with the mass $\frac{m}{3}$ and
$\frac{m}{6}$ respectively. The fermions of the first supermultiplet have
$\gamma^{12349}$ eigenvalue of 1 while those of the second have the eigenvalue
of $-1$.

One can also consider the open strings with a suitable boundary conditions
at the end of the open strings. One can use the same form of the action $(\ref{f1})$
except for the interval for $\sigma$ runs from 0 to $ \pi$ while for closed
string $\sigma$ runs from 0 to $2\pi$.
Possible supersymmetric boundary conditions or possible supersymmetric
D-brane configurations were considered in \cite{SPH},\cite{SSY} and \cite{Me}.
Corresponding boundary states were constructed in \cite{SSY} and \cite{Me}.
For the spacelike branes, the possible type of supersymmetric
D-brane configurations were tabulated in \cite{SSY}.
D0,D2,D4,D6 spacelike branes are supersymmetric and one has to
choose particular Neumann directions to satisfy the supersymmetric conditions.
We first carry out the string theory calculation in the open string channel
to get a correctly normalized amplitude and extract the contribution from
the lowest closed string modes.
We then match these results with the supergravity calculation. 
As explained in \cite{Gaberdiel},  in the standard light cone gauge
$X^{\pm}=\ap p^{+} \tau $, $X^{\pm}$ are automatically Neumann directions.
Thus {\it we will use the nonstandard light cone gauge for the open string
$X^+ = \frac{r^+}{\pi} \sigma $ 
where $r^+$ is the brane separation along the $X^+$ coordinate and $\sigma$
is the worldsheet coordinate $\sigma \in[0,\pi]$}.
The Virasoro constraint determines $X^-$ to be a Dirichlet direction as well.
In this gauge, the mass parameter appearing in the string action is 
\beq
m=\frac{\mu r^+}{\pi}. 
\label{fi}
\eeq

The interaction energy between the branes can be written as
\bea
E & = & 2 \cdot \frac{1}{2} Tr (-1)^{Fs} {\rm l n} L_0 
\nn
& = &  \int_0^{\infty} \frac{dt}{t} Tr (-1)^{Fs} e^{-L_0  t }
\eea
where $F_s$ is the spacetime fermion number and the trace is taken over the 
open strings stretched between the branes and 
$L_0 = {\it p_- H_{lc}}$ with $\it H_{lc}$ being the light-cone Hamiltonian.

For $DpDp$ and $DpD\bar{p}$ branes separated in $r^+, r^- $ and
the transversal $x^D, x^{D'}$ directions.

\bea
E & =&  \int_0^{\infty} \frac{dt}{t} e^{-2 \pi t (\frac{2 f(x_1,x_2)}{4 \pi^2 
\ap})} (  2  \sinh \pi t \omega_0 )^{2-n_N} (  2  \sinh \pi t \omega_0')^{2-n'_N}
\nn
&\cdot& \frac{f_A^{\omega_0}(q)^4f_A^{\omega'_0}(q)^4}{f_1^{ \omega_0}(q)^4
f_1^{ \omega'_0}(q)^4}
\label{f3} 
\\
f(x_1,x_2) &=& ir^+ r^- + {{m_1 r^+}\over{2 \sinh m_1
    r^+}}[((\vec{x_1^D})^2 + (\vec{x_2^D})^2) 
\cosh m_1 r^+
- 2 \vec{x_1^D}\cdot\vec{x_2^D}]\nn 
&+ &  {{m_2 r^+}\over{2 \sinh m_2 r^+}}[((\vec{x_1^{D'}})^2 + (\vec{x_2^{D'}})^2) \cosh m_2 r^+
- 2 \vec{x_1^{D'}}\cdot\vec{x_2^{D'}}]  \label{f}
\eea
where $q=e^{-2 \pi t }$ and $n_N$ is the number of Neumann directions
along 1234 while $n'_N$ is the number of those along 5678 with $n_N + n'_N = p +1 $.
Also $m_1 = {\m\over3}$, $m_2 = {\m \over 6}$,  $\omega_0 = \frac{m}{3}$, $\omega '_0 = \frac{m}{6}$ while $m$ is given by
(\ref{fi}). The $i$ factor in $ir^+ r^-$ of (\ref{f}) appears since we
consider the Euclideanized IIA plane wave background\cite{Johnson}.

The value $A=1$ for $DpDp$ and $A=4$ for $DpD\bar{p}$ configurations.
\bea
f_1^{(m)}(q) & = & q^{-\Delta_m}(1-q^m)^{1/2}\prod_{n=1}^\infty
(1-q^{\sqrt{n^2+m^2}})  \nonumber  \\
f_4^{(m)}(q) & = & q^{-\Delta'_m}\prod_{n=1}^\infty
(1-q^{\sqrt{(n-\frac{1}{2})^2+m^2}})
\eea
while $\Delta_m$, $\Delta'_m$ are defined as 
\bea 
\Delta_m &=& -\frac{1}{(2\pi)^2}\sum_{p=1}^\infty\int_0^\infty ds \  
e^{-p^2s}e^{-\frac{\pi^2m^2}{s}}
\nonumber \\
\Delta'_m &=& -\frac{1}{(2\pi)^2}\sum_{p=1}^\infty(-1)^p\int_0^\infty ds\  
e^{-p^2s}e^{-\frac{\pi^2m^2}{s}}
\eea
The modular transformation properties are given by 
\begin{equation}
f_1^m(e^{-2\pi t}) = f_1^{mt}(e^{\frac{-2\pi }{t}}),   \,\,\,\,\,
f_4^m(e^{-2\pi t}) = f_2^{mt}(e^{\frac{-2\pi }{t}})
\end{equation} 
where $f_2^{(m)}(q)$ is defined by 
\begin{equation}
f_2^{(m)}(q) =  q^{-\Delta_m}(1+q^m)^{1/2}\prod_{n=1}^\infty
(1+q^{\sqrt{n^2+m^2}}).  \nonumber 
\end{equation} 
The $DpDp$ and $DpD\bar{p}$ brane configurations are aligned along the
special directions which make the resulting Dp-brane configurations supersymmetric.
One can see that in $(\ref{f3})$, half of the contribution comes from one
supermultiplet with the mass parameter ${m\over3}$ and the other half
comes from the other supermultiplet with the mass parameter ${m\over6}$.  
. The large distance behavior of (\ref{f3}) comes from the leading behavior
of the integrand for small $t$. In order to match with the field
theory calculation, where the calculation is done on the Lorentzian
spacetime, we should take account of the Wick rotation
effect. Equivalently one follows the convention of \cite{Dabholkar, skenderis}
so that one works in the Lorentzian signature for space time with a
suitable $\epsilon$ prescription, that is, one might start with 
\begin{equation}
E  =  i \int_0^{\infty} \frac{dt}{t} Tr (-1)^{Fs} e^{i(L_0+i\epsilon)  t }
\end{equation}
 Either way, one obtains
\bea
E_{Dp-Dp} &=& - 4 \pi ( 4 \pi^2 \ap)^{3-p} \sin^2 \frac{ \mu r^+}{3}
\sin^2 \frac{\mu r^+}{6} I_0^{9-p}  \label{bb}
\\
E_{Dp-D\bar{p}} &=&  - 4 \pi ( 4 \pi^2 \ap)^{3-p} \cos^2 \frac{ \mu r^+}{3}
\cos^2 \frac{\mu r^+}{6} I_0^{9-p}
\eea
where $I_0^{9-p}$ is the integrated propagator over $(p+1)$-longitudinal
directions with the separation in $r^+,r^-$ and the transverse directions for the associated propagator
$G_0(x_1,x_2)$ satisfying $\Box G_0 (x_1,x_2) = i \d(x_1-x_2)$.
This is given by
\beq
I_0^{9-p}= i {{(i\m r^+)^{3-p}}\over2} (2\pi)^{p-4}{ \tilde{f}(x_1,x_2)}^{p-3}\Gamma (3-p)
\frac{ (\frac{\mu}{3})^{2-n_N} (\frac{\mu}{6})^{2-n'_N} }
{ \sin^2 \frac{\mu r^+}
{3} \sin^2 \frac{\mu r^+}{6} }.
\eeq
where $\tilde{f}(x_1,x_2)$ is given by 
\begin{eqnarray}
\tilde{f}(x_1,x_2) &=& r^+ r^- + {{m_1 r^+}\over{2 \sin m_1
    r^+}}[((\vec{x_1^D})^2 + (\vec{x_2^D})^2) 
\cos m_1 r^+
- 2 \vec{x_1^D}\cdot\vec{x_2^D}]\nn  \\
&+ &  {{m_2 r^+}\over{2 \sin m_2 r^+}}[((\vec{x_1^{D'}})^2 + (\vec{x_2^{D'}})^2) \cos m_2 r^+
- 2 \vec{x_1^{D'}}\cdot\vec{x_2^{D'}}].   \nonumber
\end{eqnarray}
We compare this expression with that from the field theory
 calculation on the next section and compute the tension of the
 D-branes. 

Now let us concentrate on the large $t$ behavior of the expression. 
The integrand without $\frac{1}{t}$ factor is reduced to 
\begin{equation}
exp(-2\pi t [\frac{f(x_1, x_2)}{2\pi^2
  \alpha'}+\frac{\omega_0}{2}(2-n_N)
+\frac{\omega_0'}{2}(2-n_{N'})+4(-\Delta_{\omega_0}'+\Delta_{\omega_0})
+4(-\Delta_{\omega_0'}'+\Delta_{\omega_0'})])
\end{equation}
In the $\mu \rightarrow 0$ limit, the expression inside the bracket is
reduced to 
\begin{equation}
\frac{1}{2}(\frac{2i r^+ r^- +
  (\vec{x_1^D}-\vec{x_1^D})^2+(\vec{x_1^{D'}}-\vec{x_1^{D'}})^2}
{2\pi^2\alpha'}-1)
\end{equation}
which is the same expression as one would obtain for the flat space
case. Thus if the interdistance is smaller than $X_H=\sqrt{2\pi^2
  \alpha'}$ in the flat space
the amplitude of Dbrane-anti D brane develops the divergence. This is
interpreted as D-branes having a stringy halo. This originates from
the fact that the bulk of the open strings which end on them can reach
out in the transverse directions, forming a region of potential
activity of size set by $X_H$. This halo means that the D-branes can
interact
with each other before zero separation, as there is enhancement of the
physics of interaction by new light states by the entanglement of
halos,
and the cross over into the annihilation channel begin before the
branes are coincident \cite{Johnson}. If we consider the stringy-halo effect in the
IIA plane wave background, we should extract the ground state energy
of $DpDp$ states, which is given by
$\frac{\omega_0}{2}(2-n_N)+\frac{\omega_0'}{2}(2-n_{N'})$
where $n_N+n_{N'}=p+1$.
Thus the stringy halo develops when 
\begin{equation}
2f(x_1,x_2)=2\pi^2\alpha'(8(\Delta_{\omega_0}'-\Delta_{\omega_0})
+8(\Delta_{\omega_0'}'-\Delta_{\omega_0'}))
\end{equation}
where in the $\mu \rightarrow 0$ limit the left hand side is reduced
to the interdistance of $DpD{\bar p}$ branes and the right hnd side
is reduced to $2\pi^2 \alpha'$. Due to the nonstandard light cone
gauge we should choose for the spacelike branes, $\omega_0=\frac{\mu
  r^+}{3\pi}$
and $\omega_0'=\frac{\mu r^+}{6\pi}$, the right hand side has explicit
dependence on the lightcone separation $\mu r^+$. This calculation
suggests that the tachyon dynamics for the spacelike branes in the IIA
plane wave background would be different from those in the flat space
case. Thus it would be interesting to understand the tachyon
potentials for $DpD\bar{p}$ branes in detail. The corresponding study 
in the flat space case produces huge literatures\cite{Sen}.

\section{Field Theory Calculation}

We start from the type IIA supergravity action appeared in the appendix with D-brane source terms
\beq
S_p =  -T_p \int d^{p+1} x   
 \sqrt{- \tilde g} e^{{p-3 \over 4} \Phi}  +  \m_p \int A_{[p+1]}  
\label{sourceaction} 
\eeq
where $\tilde g$ stands for the induced metric  
on the worldvolume while $T_p$ and $\mu_p$ are the brane tension and the RR charge of the D-brane 
in consideration.

In the appendix, we obtain the gravity action to the quadratic order in the fluctuation around the
plane wave background using the light-cone gauge.

The resulting action is a sum of decoupled terms characterized by coefficients {\it c} and {\it M}
\beq
S_\psi = {1 \over M} \int d^{10}x  \psi^\dagger_\alpha ( \Box - 2 i \mu c \pa_-) \psi_\alpha. 
\eeq
In our case, $\psi$ is in a tensor representation of $SO(3) \times SO(4)$ and 
a contraction of tensor indices is understood. The details are explained in the appendix.
The source terms contain the contribution of $\psi_\alpha$ with $\varepsilon=\pm 1$ for
brane/anti-brane respectively.
\beq
S_{\rm source} = \int d^{10}x\ \d^{9-p} (x - x_0)\ k ( \psi_\a + \e \bar \psi_\a). 
\eeq
The contribution of such a mode to the interaction is given by 
\beq
E  = 2 M \e  k^2  \cos \m c r^+ I_0^{9-p} .
\eeq 

\subsection{D0-brane}

We take the world volume direction to be $X^i = X^1$.
From the above source term (\ref{sourceaction}), we have
\beq
{\cal L}_{\rm source}= \frac{i T_0}{2} ( h_{11} - \frac{3}{2} \phi ) \pm \mu_0 a_1 .
\eeq
Using the notation of the appendix, this can be rewritten as 
\beq
{\cal L}_{\rm source} = { {i T_0} \over 2 }( h^\perp_{11} - { 1 \over 2}\phi_0 - { 3 \over 8}( \phi_2 + \bar\phi_2)
- {1 \over 8}( \phi_0 + \bar\phi_0 ) ) \pm \m_0 { 1 \over { 4 \sqrt{2} }}(\b_{21}
+ \bar\b_{21} + \b_{41} + \bar\b_{41} )
\eeq
where the definitions of the fields are given in the appendix.
The value of {\it M,k,c} are summerized in the following table
\begin{center}
\begin{tabular}{|c||c|c|c|}\hline
$\psi$ & $h_{11}^\perp$ & $\f_0$ & $\f_2$ \\\hline
$(M,k,c)$ & $(4\k^2, { {i T_0} \over 4},0)$ & $ ({32 \over 3} \k^2, -{{i T_0} \over 8},0)$ &
$({64 \over 9} \k^2,-{{3 i T_0} \over 16},{{\m r^+} \over 3})$\\\hline
\end{tabular}
\end{center}
\begin{center}
\begin{tabular}{|c||c|c|c|}\hline
$\psi$ & $\f_6$ & $\b_{21}$ & $\b_{41}$\\\hline
$(M,k,c)$ & $(64 \k^2,-{{i T_0} \over 16},\m r^+)$ & $(16 \k^2,{{\pm \m_0} \over {4 \sqrt{2}}},
{{\m r^+} \over 3})$ & $(16 \k^2,{{\pm \m_0} \over {4 \sqrt{2}}},
{{2 \m r^+} \over 3})$\\\hline
\end{tabular}
\end{center}

The sum of the contribution is given by
\beq
E = - \k^2 I_0^9 \left[ T_0^2 + { {T_0^2} \over 2} \cos \m r^+ + 
({{T_0^2} \over 2} \mp \m_0^2 ) \cos{ {\mu r^+} \over 3} \mp \m_0^2 \cos{ {2 \mu r^+} \over 3} \right]
\eeq
comparing with the string result, we get
\begin{center}
\beq
T_0^2=\m_0^2= {\pi \over {\k^2}} ( 4 \pi^2 \a ' )^3.
\eeq
\end{center}
Since we have $SO(3) \times SO(4)$ symmetry, the same result would be obtained if D0-brane
world voulume direction is $X^i$ with i=1,2,3.
This exhausts all the possible of the supersymmetry D-brane configurations. In the string theory
analysis, we have the precisely the same condition for the supersymmetric D0-brane.

\subsection{D2 brane}

According to the analysis in the string theory\cite{shin}, the worldvolume directions of the
supersymmetric D2-brane can be taken to be $(124)$ or $(456)$ directions.
The other configurations are equivalent to either of these by $SO(3) \times SO(4)$ rotations.
If we take the world volume direction to be $(124)$,
\bea
{\cal L}_{\rm source} & = & { {i T_2} \over 2} ( h_{11} + h_{22} + h_{44} - {1 \over 2}\phi)
\pm \m_2 a_{412}
\nn
&=& { {i T_2} \over 2} ( h^\perp_{11} + h^\perp_{22} + { 1 \over 2}\phi_0 - { 3 \over 8}( \phi_2 + \bar\phi_2)
- {1 \over 8}( \phi_0 + \bar\phi_0 ) ) 
\nn
& &
\pm \m_2 { 1 \over { 4 \sqrt{2} }}(\b_{23}
+ \bar\b_{23} - \b_{43} - \bar\b_{43} )
\eea
Again the contribution of the each field to the interaction energy is given by the table below. 
\begin{center}
\begin{tabular}{|c||c|c|c|}\hline 
$\psi$& $ h_{11}^\perp \;, h_{22}^\perp $ & $\phi_0$ & $ \f_2$\\\hline 
$(M,k,c)$ & $(4\k^2,{{iT_2} \over 4},0)$ & $({ 32 \over 3} \k^2,{{iT_2} \over 8},0)$ 
& $({64\over 9}\k^2,-{{i3 T_2}\over 16},{{\m r^+} \over 3 })$  
\\\hline 
\end{tabular} 
\end{center}

\begin{center}
\begin{tabular}{|c||c|c|c|}\hline
$\psi$ & $\f_6$ &$\b_{23}$&$\b_{43}$\\\hline
$(M,k,c)$ & $(64 \k^2,-{{i T_2} \over 16}, \m r^+)$ & $(16 \k^2,{{\pm \m_2} \over { 4 \sqrt{2}}},
{{\mu r^+} \over 3})$ & $(16 \k^2,{{\pm \m_2} \over { 4 \sqrt{2}}},
{{2 \mu r^+} \over 3})$\\\hline  
\end{tabular}
\end{center}

The interaction energy is given by
\beq
E = - \k^2 I_0^7 \left[ T_2^2 + { {T_2^2} \over 2} \cos \m r^+ + 
({{T_2^2} \over 2} \mp \m_2^2 ) \cos{ {\mu r^+} \over 3} \mp \m_2^2 \cos{ {2 \mu r^+} \over 3} \right]
\eeq
Again comparing with the string theory result, we get
\begin{center}
\beq
T_2^2 = \m_2^2 = { { 4 \pi^2 \a ' } \over {\k^2}}
\eeq
\end{center}
When the world volume direction of D2-brane is 456, one can carry out similar computation, which
gives the same values of $T_2$ and $\m_2$.  

\subsection{D4 brane}

If we take the world volume direction of the D4-brane to be $(35678)$, then the coupling
\beq
\m_4 \int A_{[5]} = \m_4 \int d^5 x \  A_{35678} = - \m_4 \int d^5 x \  \varepsilon^{124}\ _{35678} A_{124}
\label{D4source}
\eeq
in the light cone gauge where $\varepsilon_{i_1i_2\cdot\cdot\cdot i_8}$ is the totally antisymmetric
tensor of rank 8.
Thus the D4-brane in consideration is magnetically charged to the 3-form potential but
for the computational purpose the contribution from (\ref{D4source}) would be the same as
in the D2-brane case if we replace $\m_2$ by $\m_4$.
Thus we have to figure out the contribution from other remaining source terms.
\bea
{\cal L'}_{\rm source} & = & { {i T_4} \over 2} ( h_{33} + h_{55} + h_{66} + h_{77} + h_{88}
+ {\f \over 2} )\nn
& = & {{i T_4} \over 2} ( h^\perp_{33} + h^\perp_{55} + h^\perp_{66} + h^\perp_{77} + h^\perp_{88} 
\nn
& & - { 1 \over 2}\f_0 +{1 \over 8} ( \f_6 + \bar\f_6 ) + {3 \over 8} ( \f_2 + \bar\f_2 ) )
\eea
The contribution of $\f_0,\f_2,\f_6$ are already worked out in D2-brane case. Working out
the contribution from the remaining gravitational sector we obtain the interaction energy
\beq
E = - \k^2 I_0^5 \left[ T_4^2 + { {T_4^2} \over 2} \cos \m r^+ + 
({{T_4^2} \over 2} \mp \m_4^2 ) \cos{ {\mu r^+} \over 3} \mp \m_4^2 \cos{ {2 \mu r^+} \over 3} \right].
\eeq 
By the comparison with the string theory result, we get
\beq
T^2_4 = \m_4^2 = {\pi \over {\k^2}} ( 4 \pi^2 \a')^{-1}.
\eeq
If the world volume direction is $(12378)$, we can work out the similar computation, which
gives the same value of $T_4$ and $\m_4$.
All the other supersymmetric D4-brane configurations are again related to the above two cases by
$SO(3) \times SO(4)$ rotations.

\subsection{D6 brane}

All supersymmetric configurations are equivalent to the D6-brane with world volume
$(1245678)$. Again
\beq 
\m_6 \int [A_6] = \m_6 \int d^6 x \  a_3
\eeq
and we can borrow the computational result from the D0-brane.
The other source terms are
\bea
{\cal L'}_{\rm source} &=& {{i T_6} \over 2} ( \sum_{i=1245678} h_{ii} + {3 \over 2}\f ) \nn
&=& {{i T_6} \over 2} ( \sum_{i=125678} h^\perp_{ii} + {1 \over 2} \f_0
+ {3 \over 8}(\f_2 + \bar\f_2) + {1\over 8}(\f_6 + \bar\f_6)).
\eea
By the similar calculation, one can obtain the interaction energy
\beq
E = - \k^2 I_0^3 \left[ T_6^2 + { {T_6^2} \over 2} \cos \m r^+ + 
({{T_6^2} \over 2} \mp \m_6^2 ) \cos{ {\mu r^+} \over 3} \mp \m_6^2 \cos{ {2 \mu r^+} \over 3} \right].
\eeq 
Thus 
\beq
T_6^2 = \m_6^2 = {\pi \over{\k^2}}(4 \pi^2 \a')^{-3}
\eeq to match the string theory computation.
In conclusion, we find that for all supersymmetric Dp-branes of interest their
tensions and charges are given by
\beq
T_p^2 = \m_p^2 = {\pi \over \k^2}(4 \pi^2 \a')^{3-p}.
\eeq 

\section{Discussion of integrated amplitudes}
In \cite{skenderis}, the detailed study of interaction energy as functions of the brane 
separation in the IIB wave was presented. This is the continuation of their work on the 
branes and strings on the Type IIB plane wave background and AdS space\cite{st1, st2,st3}.
Also see a related work\cite{gg}.
 Many special properties of the amplitude 
are traced to geometric properties of the plane wave background with emphasis on  
 the lack of translational invariance. For special values of $r^+$, 
the translational invariance is restored, which is related to the focusing of the 
geodesics. In the IIA case, many of the discussions can be carried over. 
Many features of the amplitude could again be traced to the lack of translational
invariance and for special values of $r^+$, the translational invariance is 
restored. In contrast to the IIB case, for some other special values of $r^+$, 
the translational invariance is partially restored. This is related to the fact that 
the worldsheet theory of the Type IIA plane wave background 
consists of two (4,4) supermultiplets, 
one has mass $\frac{m}{3}$ and the other has 
mass $\frac{m}{6}$ while Type IIB theory correspond to two (4,4) supermultiplets with the same mass
$m$, which leads to the (8,8) worldsheet supersymmetry. 

In the Type IIA plane wave, the translation in the $x^I=(x^i, x^{i'})$ directions acts on 
\begin{eqnarray}
\delta r^- &=& -{\rm sin}\frac{\mu}{3}x^+ \epsilon^i x^i 
-{\rm sin}\frac{\mu}{6}x^+ \epsilon^{i'} x^{i'} \label{ta}\\
\delta x^i &=& {\rm cos} \frac{\mu}{3} x^+ \epsilon ^i  \nonumber \\
 \delta x^{i'} &=& {\rm cos} \frac{\mu}{6} x^+ \epsilon^{i'}  \nonumber
\end{eqnarray}
where $i$ runs from 1 to 4 and $i'$ runs from 5 to 8. 
This implies that the system of branes separated along the lightlike directions is not 
geometrically invariant under the translations in the $x^I$ directions. 
The integrated propagators have very different behaviors than that of the flat case.
The massless propagator $G_0(x_1, x_2)$ satisfying $\Box G_0(x_1, x_2)=i \delta(x_1-x_2)$
is given by 
\begin{equation}
G_0(x_1, x_2)= i (\frac{\frac{\mu r^+}{3}}{\sin \frac{\mu r^+}{3}})^2
(\frac{\frac{\mu r^+}{6}}{\sin \frac{\mu r^+}{6}})^2 \int_0^{\infty} 
\frac{ds}{(4\pi i s)^5} exp(-\frac{\sigma+i\epsilon}{4is})  \label{green}
\end{equation}
where 
\begin{eqnarray}
\sigma(x^i, x^{i'})&=&2r^+r^- 
+ \frac{\frac{\mu r^+}{3}}{\sin \frac{\mu r^+}{3}}
 \Sigma_{i=1}^4[(x_1^ix_1^i + x_2^i x_2^i)\cos \frac{\mu r^+}{3}-2x_1^ix_2^i]  \\
 &+& \frac{\frac{\mu r^+}{6}}{\sin \frac{\mu r^+}{6}}
\Sigma_{i'=5}^8[(x_1^{i'}x_1^{i'} + x_2^{i'} x_2^{i'})\cos \frac{\mu r^+}{6}-2x_1^{i'}x_2^{i'}]. 
\nonumber
\end{eqnarray}
Compared with the Type IIB case we observe the splitting of 
\begin{equation}
(\frac{\mu r^+}{{\rm sin} \mu r^+})^4
\end{equation}
into 
\begin{equation}
(\frac{\frac{\mu r^+}{3}}{{\rm sin} \frac{\mu r^+}{3}})^2
(\frac{\frac{\mu r^+}{6}}{{\rm sin} \frac{\mu r^+}{6}})^2
\end{equation}
in the corresponding expression in the Type IIB case.
 This is related to the fact that the 8 transverse components of Type IIB theory
involves the harmonic oscillator eigenfunctions with the parameter $\mu r^+$, while in Type IIA
theory 4 components have those with $\frac{\mu r^+}{3}$ and the other components have those with 
$\frac{\mu r^+}{6}$. 
If we take the flat space limit $\mu \rightarrow 0$ 
we have a translational invariance along $x^i$ and $x^{i'}$ directions and 
$G_0(x_1, x_2)$ of (\ref{green}) is reduced to the propagator in the flat space. 
If we integrate the $p+1$ dimensional world volume directions in the 
$\mu \rightarrow 0$ limit,  we obtain the amplitude in the flat space
\begin{equation}
Z_{flat} \sim V_{p+1}(\sigma^2_{(9-p)})^{-\frac{7-p}{2}}
\end{equation}
but while in the IIA case
\begin{equation}
Z_p \sim \int_0^{\infty} \frac{ds}{s^{4-p}}
exp(-\frac{\sigma_{9-p}}{i s})
\end{equation}
which gives the brane/brane exchange behavior of (\ref{bb}). 
We don't have any volume factors and the power law of the amplitude in the flat space 
is modified. 

This behavior also can be seen from that of supergravity fields. 
The long range supergravity field sourced by the brane depends on the position of the 
Neumann directions. If we look for the behavior of a massless supergravity mode $\Psi$
($c=0$) at a given point far from the brane, one gets 
\begin{equation}
\Psi \sim (\frac{\frac{\mu r^+}{3}}{{\rm sin} \frac{\mu r^+}{3}})^2
(\frac{\frac{\mu r^+}{6}}{{\rm sin} \frac{\mu r^+}{6}})^2 
\,\,{\rm tan}^{\frac{n_1}{2}}\frac{\mu r^+}{3}
{\rm tan}^{\frac{n_2}{2}}\frac{\mu r^+}{6} \tilde{\sigma}^{n_1+n_2-8}
\end{equation}
where $n_1$ and $n_2$ is the number of Neumann directions in $x^i$ directions and in 
$x^{i'}$ directions respectively with $n_1+n_2=p+1$ and 
\begin{equation}
\tilde{\sigma}=-\frac{\mu r^+}{3}{\rm tan}\frac{\mu r^+}{3}\Sigma_{r=1}^{n_1} (x^r)^2
-\frac{\mu r^+}{6}{\rm tan}\frac{\mu r^+}{6}\Sigma_{r'=5}^{n_2+4} (x^{r'})^2 
+\sigma(x^r=0, x^{r'}=0).
\end{equation}
As happens in the IIB case, the powerlaw dependence is the same as in the flat space
but the field sourced is not translationally invariant along the directions parallel
to the brane. Integrating with respect to $x^r, x^{r'}$ of Neumann directions 
gives the brane/brane exchange 
behavior.

However, for special values of the separation, the translational invariance is restored.
If $\mu r^+=6n\pi$ with $n$ being an integer, the translation action (\ref{ta}) is given by
\begin{equation}
\delta x^-= 0, \,\,\, \delta x^i= \epsilon ^i  \,\,\,
\delta x^{i'}=(-1)^n \epsilon^{i'}
\end{equation}
while if $\mu r^+=3(2n+1)\pi$
\begin{equation}
\delta x^-=\frac{\mu}{6}(-1)^{n+1}\epsilon^{i'}x^{i'}  \,\,\,
\delta x^i=-\epsilon^i  \,\,\,  \delta x^{i'}=0.
\end{equation}
In this case we have the translational invariance along $x^i$ directions but not along 
$x^{i'}$ directions. Depending on the value of $\mu r^+$ one can have the full restoration 
of translational invariance or the partial restoration of the translational invariance. 
The special value of $\mu r^+$ could be understood from the behavior of the geodesics as well.
The geodesics in the Type IIA plane wave background is given by 
\begin{eqnarray}
r^- &=& \tilde{c} r^+ + \frac{\mu}{3}\Sigma_{i=1}^4[(\frac{1}{4}((p_0^i)^2-(x_0^i)^2){\rm sin} 
\frac{2\mu r^+}{3}
+\frac{1}{2}x_0^i p_0^i {\rm cos}\frac{2\mu r^+}{3}] \\
    &+ &  \frac{\mu}{6}\Sigma_{i'=5}^8[\frac{1}{4}((p_0^{i'})^2-(x_0^{i'})^2){\rm sin} 
\frac{\mu r^+}{3}
+\frac{1}{2}x_0^{i'} p_0^{i'} {\rm cos}\frac{\mu r^+}{3}] \nonumber \\
x^i &=& x_0^i {\rm cos}\frac{\mu r^+}{3} +p_0^i {\rm sin} \frac{\mu r^+}{3}  \nonumber  \\
x^{i'} &=& x_0^{i'}  {\rm cos}\frac{\mu r^+}{6} +p_0^{i'} {\rm sin} \frac{\mu r^+}{6} 
\end{eqnarray}
if we parametrize the geodesics by $r^+$.
A generic geodesic will reconverge to its original transverse position after 
$\mu r^+=12 n\pi$. Note that for $\mu r^+=6n\pi$
\begin{equation}
x^{i'}=(-1)^n x_0^{i'},  \,\,\, x^i=x_0^i.
\end{equation}
For $\mu r^+=6n\pi$ the field theory result has the reinstated translational
invariance. 
If $x_1^i\neq x_2^i$ and $x_1^{i'}\neq (-1)^n x_2^{i'}$, 
 the geodesics will be of infinite distance
and we regularize the geodesic distance 
\begin{equation}
\mu r^+= 6n\pi +\epsilon
\end{equation}
then 
\begin{equation}
\sigma_{10}=(\frac{6n\pi}{\epsilon}+1)[ \Sigma_i(x_1^i-x_2^i)^2
+\Sigma_{i'}(x_1^{i'}-(-1)^n x_2^{i'})^2]
+2r^+ r^- +o(\epsilon).
\end{equation}
This expression for the distance is very similar to that of the flat space.
In this limit the field theory exchange expression is given by 
\begin{eqnarray}
Z_p&\sim& V_{n_1}V_{n_2}\epsilon^{\frac{n_1+n_2}{2}}
\int \frac{ds}{s^{5-\frac{n_1}{2}-\frac{n_2}{2}}}exp\frac{i\sigma_{(10-n_1-n_2)}}{s} \\
  &\sim & \frac{V_{n_1}V_{n_2}\epsilon^4}
{[\Sigma_{r=n_1+1}^4(x_1^r-x_2^r)^2
+\Sigma_{r'=n_2+5}^8(x_1^{r'}-(-1)^n x_2^{r'})^2]^{4-\frac{n_1}{2}-\frac{n_2}{2}}} \nonumber
\end{eqnarray}
,which shows the reinstated translational invariance.

If $x_1^r= x_2^r$ for $j_1$ directions among $x^is$ and $x_1^{r'}= (-1)^n x_2^{r'}$ for
$j_2$ directions among $x^{i'}s$ then $Z_P$ is given by
\begin{equation}
Z_p \sim  \frac{V_{n_1}V_{n_2}\hat{V}_{j_1}\hat{V}_{j_2}\epsilon^4}
{[\Sigma_{r=j_1+1}^4(x_1^r-x_2^r)^2+\Sigma_{r'=j_2+5}^8(x_1^{r'}-(-1)^n
  x_2^{r'})^2]^{4-\frac{n_1+n_2}{2}-\frac{j_1+j_2}{2}}}
\end{equation}
Here $\hat{V}_j$ is the regulated momentum space volume for the
$j_1,j_2$ directions.\footnote{We use $\delta(x)=\lim_{\alpha\rightarrow \infty}
  \frac{\alpha}{\sqrt{\pi}}exp(-\alpha^2 x^2)=\frac{\hat{V}}{2\pi}$
  with $\hat{V}$ being the volume of the momentum space.}
If $\mu r^+=3(2n+1)\pi+\epsilon$ 
\begin{equation}
\sigma_{10}=2r^+r^- -\frac{3(2n+1)\pi}{\epsilon}[\Sigma_{i=1}^4(x_1^i+x_2^i)^2]
+(-1)^{n+1}\frac{(2n+1)\pi}{2}[\Sigma_{i'=5}^8 2x_1^{i'}x_2^{i'}] +o(\epsilon)
\end{equation}
The translational invariance is restored only in the $x^i$ directions.
If $x_1^i\neq -x_2^i$ 
\begin{eqnarray}
Z_p&\sim& V_{n_1}\epsilon^{\frac{n_1}{2}}
\int \frac{ds}{s^{5-\frac{n_1}{2}}}exp\frac{i\sigma_{(10-n_1-n_2)}}{s} \\
  &\sim & \frac{V_{n_1}\epsilon^4}
{[\Sigma_{r=n_1+1}^4(x_1^r-x_2^r)^2]^{4-\frac{n_1}{2}}}.
\end{eqnarray}
If $x_1^i=-x_2^i$ for  $j_1$ directions
\begin{equation}
Z_p \sim \frac{V_{n_1}\hat{V}_{j_1}\epsilon^4 }
{\Sigma_{i=n_1+j_1+1}^4((x_1^i+x_2^i)^2)^{4-\frac{n_1+j_1}{2}}}
\end{equation}
and if $x_1^i=-x_2^i$ for all $i$
we have
\begin{equation}
Z_p \sim V_{n_1}\hat{V}_{4-n_1}\epsilon^2
\int \frac{ds}{s^3}exp\frac{i}{s}(2r^+r^-
+(-1)^{n+1}\frac{(2n+1)\pi}{2}\Sigma_{i'=n_1+5}^8 2x^{i'}x^{i'})
\end{equation}
In all the cases where the translation invariance is at least partially restored, 
the amplitude vanishes in the limit $\epsilon
\rightarrow 0$, whose feature is shared by the exchange amplitudes of the supersymmetric 
D-branes in the 
flat space.

As happens in the Type IIB case, the special features are also observed in the 
worldsheet calculation. We follow the convention in \cite{HS, HS2, Me}.
 If $\mu r^+ \rightarrow 3\pi$, the bosonic mode expansion 
in $X^i$ directions develop new zero modes 
\begin{eqnarray}
X^{i_N}&=& (x^{i_N}+\alpha' p^{i_N}\tau) 2\cos\sigma +\cdots\\
X^{i_D}&=&x_0^{i_D}\cos\sigma + (x^{i_D}+\alpha' p^{i_D}\tau)2\sin\sigma +\cdots
\nonumber
\end{eqnarray}
with $[x^i, p^j]=i\delta^{ij}$. 
The corresponding fermionic zero modes are given by
\begin{eqnarray}
\Psi_2^+ &=& \sqrt{\frac{\alpha'}{2}} (\Psi_1
e^{-i\sigma}-i\gamma^4\Omega \Psi_1e^{i\sigma}+\Psi_{-1}e^{i\sigma} 
+i\gamma_4\Omega \Psi_{-1}e^{-i\sigma}) +\cdots \\
\Psi_1^- &=&  \sqrt{\frac{\alpha'}{2}} (\Omega\Psi_1
e^{i\sigma}-i\gamma^4\Psi_1e^{-i\sigma}+\Omega\Psi_{-1}e^{-i\sigma} 
+i\gamma_4\Psi_{-1}e^{i\sigma}) +\cdots \nonumber
\end{eqnarray}
with $\{\Psi_1^a, \Psi_{-1}^b\}=\delta^{ab}$ where $a,b$ runs from 1 to
4. $\Omega$ is the matrix relating fermionic left and right modes in the open string sectors
whose detailed forms are dependent on the D-branes of interest and are given in \cite{SSY, Me}.
 No new zero bosonic modes in $X^{i'}$ directions appear and neither 
the fermionic modes corresponding to their superpartners do. The interaction energy 
for the brane/brane in this case is given by 
\begin{equation}
E=V_{n_1}V_{4-n_1}\int_0^{\infty}\frac{dt}{t}exp(\frac{3ir^-}{\alpha' \mu}t)
(2{\rm sinh}\pi t)^{2-n_1}(2{\rm sinh} \frac{\pi t}{2})^{2-n_2} (1-1)^4
\end{equation}
where $V_{n_1}$ is the volume of the Neumann directions within $x^i$ directions 
and $V_{4-n_1}$ is the volume of the remaining Dirichlet directions 
in the $x^i$ directions and $(1-1)^4$ comes from the fermionic zero modes, which annihilates
the amplitude.
Existence of continuos modes in the Neumann directions reflects the partially reinstated 
translational invariance along the world volume directions within $x^i$ directions
leading to $V_{n_1}$ factors. There are also continuous modes in the Dirichlet directions
leading to $V_{4-n_1}$ factor. This follows from the infinite family of geodesics connecting 
$x_0^{i_D}$ to $-x_0^{i_D}$.
 
If $\mu r^+ \rightarrow 6\pi$ all of the bosonic and the fermionic modes develop new zero modes.
The bosonic zero  modes are given by 
\begin{eqnarray}
X^{i_N'}&=& (x^{i_N'}+\alpha' p^{i_N'}\tau) 2{\rm cos}\sigma +\cdots \label{bz}\\
X^{i_D'}&=&x_0^{i_D'}{\rm cos}\sigma + (x^{i_D'}+\alpha' p^{i_D'}\tau)2{\rm sin}\sigma +\cdots
\nonumber
\end{eqnarray}
and the fermionic zero modes are given by 
\begin{eqnarray}
\Psi_2^- &=& \sqrt{\frac{\alpha'}{2}} (\Psi_1'
e^{-i\sigma}+i\gamma^4\Omega \Psi_1'e^{i\sigma}+\Psi_{-1}'e^{i\sigma}
-i\gamma_4\Omega \Psi_{-1}'e^{-i\sigma}) +\cdots \label{fz} \\
\Psi_1^+ &=&  \sqrt{\frac{\alpha'}{2}} (\Omega\Psi_1'
e^{i\sigma}+i\gamma^4\Psi_1'e^{-i\sigma}+\Omega\Psi_{-1}'e^{-i\sigma}
-i\gamma_4\Omega\Psi_{-1}'e^{i\sigma})  +\cdots \nonumber
\end{eqnarray}
while for $x^i$ directions we just replace $\sigma$ by $2\sigma$
for the bosonic zero modes and the fermionic zero modes in (\ref{bz}) and ({\ref{fz}). 
The interaction energy is given by 
\begin{equation}
E=V_{n_1}V_{4-n_1}V_{n_2}V_{4-n_2}
\int_0^{\infty}\frac{dt}{t}exp(\frac{6ir^-}{\alpha' \mu}t)
(2{\rm sinh} 2\pi t)^{2-n_1}(2{\rm sinh} \pi t)^{2-n_2} (1-1)^8
\end{equation}
where $(1-1)^8$ part comes from the fermionic zero modes, $V_{n_2}$ is the volume
of the Neumann directions within $x^{i'}$ directions and $V_{4-n_2}$ is the volume 
of the remaining Dirichlet directions within $x^{i'}$ directions.

\begin{appendix}
\section{Appendix}
In this appendix, we explain how to obtain the bulk action of the relevant field
for the computation in the main text.
In \cite{shin}, the equations of motion for the IIA supergravity fluctuation to the
leading order are obtained. They consists of the decoupled equation of motion of the form
\beq 
\Box \f_c + i { 1\over 3} \  \m c \pa_- \f_c = 0
\label{master}
\eeq
We use the leftmost subscript to indicate the coefficient {\it c} in the
above equation in writing the various fields.
The bosonic fields of IIA supergravity theory consist of the dilaton $\Phi$,
graviton $g_{\m \n}$, one-form field $A_\m$, two form field $B_{\m \n}$ and
three form field potential $A_{\m \n \rho}$.
We consider the small fluctuation of the above fields on the IIA plane wave background.
\bea
\Phi &=& \phi \qquad
g_{\m \n} = \bar g_{\m \n} + h_{\m \n}\nn
A_\mu &=& \bar A_\mu + a_\mu \qquad
B_{\mu\nu} = b_{\mu\nu} 
\nonumber \\
A_{\mu\nu\rho} &=& \bar A_{\mu\nu\rho} + a_{\mu\nu\rho}
\eea
where $\bar g_{\m \nu}$, $\bar A_\m$ and $\bar A_{\mu\nu\rho}$ denote the IIA plane-wave background.
We take the light cone gauge.
\beq
a_- = b_{-I} = a_{-IJ} = h_{-I} = 0
\eeq
Since the IIA plane-wave background has $SO(3) \times SO(4)$ symmetry,
it's convenient to classify the small fluctuations using this symmetry.
$SO(3) \times SO(4)$ scalar multiplets are given by
\bea
\phi_0 &\equiv&  \phi + \frac{1}{3} h_{ii} + h_{44} ~,
  \nonumber \\
\phi_2 &\equiv& \phi + \frac{4}{3} i a_4 - \frac{2}{3} h_{44} ~,~~~
\bar\phi_2 \equiv \phi - \frac{4}{3} i a_4 - \frac{2}{3} h_{44} ~,
  \nonumber \\
\phi_6 &\equiv& \phi - 4 i a_{123} - 2 h_{ii} ~,~~~
    \bar\phi_6 \equiv \phi + 4 i a_{123} - 2 h_{ii} ~.
\eea
Here the subscript indicates the coefficient appearing in the mass parameter in the
equation of motion (\ref{master}). In the appendix, {\it i} denotes $SO(3)$ direction while
{\it i'} denotes $SO(4)$ one.
For example, $\f_6$ satisfies 
\beq 
\Box \f_6 + i 2  \m  \pa_- \f_6 = 0
\eeq
and its complex conjugate $\bar \f_6$ satisfies
\beq
\Box \bar \f_6 - i 2  \m  \pa_- \bar \f_6 = 0.
\eeq
$SO(3)$ vector multiplets are
\bea
\beta_{0i} &\equiv& b_{4i} \\
\beta_{2i} &\equiv& \sqrt{2}a_i - i h_{4i} + \frac{i}{2}\epsilon_{ijk}
b_{jk}
               + \sqrt{2}\epsilon_{ijk}a_{4jk},
\quad \bar\beta_{2i} \equiv \sqrt{2}a_i + i h_{4i} -
\frac{i}{2}\epsilon_{ijk} b_{jk}
                + \sqrt{2}\epsilon_{ijk}a_{4jk},
\nonumber \\
\beta_{4i} &\equiv& \sqrt{2}a_i + i h_{4i} + \frac{i}{2}\epsilon_{ijk}
b_{jk}
                - \sqrt{2}\epsilon_{ijk}a_{4jk},
\quad \bar\beta_{4i} \equiv \sqrt{2}a_i - i h_{4i} -
\frac{i}{2}\epsilon_{ijk} b_{jk}
               - \sqrt{2}\epsilon_{ijk}a_{4jk} \nonumber
\eea
Here again the leftmost subscript denotes the coefficient in the mass parameter. 
$SO(4)$ vector multiplets are given by
\bea
&&\beta_{1i'} \equiv \sqrt{2}a_{i'} +  i h_{4i'},\qquad \bar\beta_{1i'}
\equiv \sqrt{2}a_{i'} - i h_{4i'},
\nn
&& \beta_{3i'} \equiv b_{4i'}
                    - \frac{i}{3} \epsilon_{i'j'k'l'} a_{j'k'l'},
\qquad \bar \beta_{3i'} \equiv b_{4i'}
                    + \frac{i}{3} \epsilon_{i'j'k'l'} a_{j'k'l'}
\eea
Then there are two index tensor fields.
$(ij)$ and $(i'j')$ components are $SO(3)$ and $SO(4)$ graviton multiplets respectively.
Thus we have
\bea
&&h^\perp_{ij} \equiv h_{ij} - {1 \over 3} \d_{ij} h_{kk}, \qquad
h^\perp_{i'j'} \equiv h_{i'j'} - {1 \over 4} \d_{i'j'} h_{k'k'} 
\eea
and the equations of motion are written as
\bea
&&\Box h^\perp_{ij} = 0, \qquad
\Box h^\perp_{i'j'} = 0
\eea
And $(ij')$ components are defined by
\bea
\beta_{1ij'} \equiv b_{ij'} + i \sqrt{2}a_{4ij'}, \qquad \bar\beta_{1ij'}
\equiv b_{ij'} - i \sqrt{2}a_{4ij'},
\eea 
and $(i'j')$ components are
\bea
\beta_{2i'j'} &\equiv& a^+_{i'j'} - i {\tilde a^+}_{i'j'}, \qquad
\bar\beta_{2i'j'} \equiv a^+_{i'j'} + i {\tilde a^+}_{i'j'},
\nonumber \\
\beta_{4i'j'} &\equiv& a^-_{i'j'} + i {\tilde a^-}_{i'j'}, \qquad
\bar\beta_{4i'j'} \equiv a^-_{i'j'} - i {\tilde a^-}_{i'j'},
\eea
where
\bea
a^{\pm}_{i'j'} \equiv b_{i'j'}
                           \pm \frac{1}{2}\epsilon_{i'j'k'l'}
                            b_{k'l'},
\qquad { 1\over\sqrt{2}}{\tilde a^{\pm}}_{i'j'} \equiv  a_{4i'j'}
                                \pm \frac{1}{2}\epsilon_{i'j'k'l'}
                                a_{4k'l'}.
\nonumber
\eea 
Finally we have two types of 3-rank tensor fields.
$(ijk')$ components are
\bea
\beta_{3ijk'} \equiv \sqrt{{2\over 3}}a_{ijk'} - i \epsilon_{ijk} h_{kk'}, \qquad
\bar\beta_{3ijk'} \equiv \sqrt{{2\over 3}}a_{ijk'} + i \epsilon_{ijk} h_{kk'}, 
\eea
while $(ij'k')$ components are
\beq
\b_{0ij'k'} \equiv a_{ij'k'}.
\eeq
Consider the above diagonalized fields. One can obtain the action which
gives the equation fo motion of each field.
Since we started with the equation of motion, there are same
ambiguities in the relative normalization of the fields. However
such relative normalizations could be fixed by matching the 
gravity and the 2-form potential parts with the known form of the IIA actions in the light-cone
gauge. 

The relevant part of the action needed for the computation in the
main text is given by 
\bea
{\cal L} &=& {1 \over {8\k^2}} ( h^\perp_{ij} \Box h^\perp_{ij} + h^\perp_{i'j'} \Box h^\perp_{i'j'} ) \nn
&+& {3\over{32\k^2}}\f_0 \Box \f_0 + {9\over{64\k^2}}\bar\f_2( \Box + {{2 i \m}\over3} \pa_- ) \f_2
+ {1\over{64\k^2}}\bar\f_6 ( \Box + 2 i \m \pa_- ) \f_6 \nn
&+& {1\over{16\k^2}}\bar\b_{2i} ( \Box + {{2 i \m}\over3} \pa_- ) \b_{2i} +
{1\over{16\k^2}}\bar\b_{4i} ( \Box + {{4 i \m}\over3} \pa_- ) \b_{4i} \nn
&+& {1\over{8\k^2}}\bar\b_{1i'}( \Box + {{ i \m}\over3} \pa_- ) \b_{1i'} +
{1\over{16\k^2}}\bar\b_{3i'}( \Box +  i \m \pa_- ) \b_{3i'} \nn
&+& {1\over{16\k^2}}\bar\b_{1ij'}( \Box + {{ i \m}\over3} \pa_- ) \b_{1ij'}\nn
&+& {1\over{64\k^2}}\bar\b_{2i'j'}( \Box + {{ 2 i \m}\over3} \pa_- ) \b_{2i'j'} +
{1\over{64\k^2}}\bar\b_{4i'j'}( \Box + {{ 4 i \m}\over3} \pa_- ) \b_{4i'j'} \nn
&+& {1\over{16\k^2}}\bar\b_{3ijk'}( \Box + i \m \pa_- ) \b_{3ijk'} \nn
&+& \cdot \cdot \cdot
\eea
where the omitted terms are not relevant for the computation in the main text.
There are slight differences in the field definitions adopted here and in \cite{shin}.
Our convention is that
the coefficient in front of each field is consistent with the usual normalization of the p -form field
\beq
\frac{1}{p!}\frac{1}{4 \k^2} \int d^{10} x \  A_{[i_1 \cdot\cdot\cdot i_p]} \Box A_{[i_1 \cdot\cdot\cdot i_p]}
= \frac{1}{4 \k^2} \int d^{10} x \  \sum_{i_1 < i_2 < \cdot\cdot\cdot < i_p} A_{i_1 i_2 \cdot\cdot\cdot i_p}
\Box A_{i_1 i_2 \cdot\cdot\cdot i_p}.
\eeq
The integrated propagator $I^{9-p}_c (r^+,r^-)$ associated with (\ref{master}) is
simply $e^{{i c \m r^+}\over3} I^{9-p}_0 (r^+,r^-)$.

\end{appendix}

\vs{.5cm}
\noindent
{\large\bf Acknowledgments}\\[.2cm]
This work  is supported by the 
Korea Science and Engineering Foundation(KOSEF) Grant R01-2004-000-10526-0 (JP, YK)
and by the Science Research Center Program of KOSEF through the Center for 
Quantum Spacetime(CQUeST) of Sogang University with grant number R11-2005-021 (JP). 


\newcommand{\Jn}[4]{{\sl #1} {\bf #2} (#3) #4}
\newcommand{\andJ}[3]{{\bf #1} (#2) #3}
\newcommand{\AP}{Ann.\ Phys.\ (N.Y.)}
\newcommand{\MPL}{Mod.\ Phys.\ Lett.}
\newcommand{\NP}{Nucl.\ Phys.}
\newcommand{\PL}{Phys.\ Lett.}
\newcommand{\PR}{Phys.\ Rev.}
\newcommand{\PRL}{Phys.\ Rev.\ Lett.}
\newcommand{\PTP}{Prog.\ Theor.\ Phys.}
\newcommand{\hepth}[1]{{\tt hep-th/#1}}

\end{document}